\documentclass{appolb}
\usepackage{amsthm}
\usepackage{amssymb}
\usepackage{graphicx,xcolor}
\usepackage[backend=bibtex, sorting=none, defernumbers, doi=false]{biblatex}
\usepackage{hyperref}
\hypersetup{
    colorlinks = true,
    linkcolor = {blue},
    citecolor = {blue},
}
\AtEveryCitekey{\clearfield{title}}
\AtEveryBibitem{\clearfield{title}}
\newcommand{\simulat}{\texttt{SIMULATeQCD}}
\newcommand \hmu {\hat{\mu}}

\begin{document}
\title{Detecting critical points from Lee-Yang edge singularities in lattice QCD%
\thanks{Presented at the 29TH INTERNATIONAL CONFERENCE ON ULTRARELATIVISTIC NUCLEUS - NUCLEUS COLLISIONS, April 4-10, 2022, Kraków, Poland.}%
}
\author{Christian Schmidt, David A. Clarke,  Guido Nicotra
\address{Fakultät für Physik, Univeristät Bielefeld, D-33615 Bielefeld, Germany}
\\[3mm]
{Francesco Di Renzo, Petros Dimopoulos, Simran Singh
\address{Dipartimento di Scienze Matematiche, Fisiche e Informatiche,\\ Università di Parma and INFN, Gruppo Collegato di Parma I-43100 Parma, Italy}
}
\\[3mm]
Jishnu Goswami
\address{RIKEN Center for Computational Science, Kobe 650-0047, Japan}
\\[3mm]
Kevin Zambello
\address{Università di Pisa and INFN Sezione di Pisa, Largo B. Pontecorvo 3, I-56127 Pisa, Italy}
}
\maketitle
\begin{abstract}
A new approach is presented to explore the singularity structure of lattice QCD in the complex chemical potential plane.
Our method can be seen as a combination of the Taylor expansion and analytic continuation approaches.
Its novelty lies in using rational (Padé) approximants for studying Lee-Yang edge singularities.
We present a calculation of the cumulants of the net-baryon number as a function of a purely imaginary baryon number chemical potential, obtained with highly improved staggered quarks at temporal lattice extent of $N_\tau=4,6$. 
We construct various rational function approximations of the lattice data and determine their poles (and roots) in the complex plane.
We compare the position of the closest pole to the theoretically expected position of the Lee-Yang edge singularity. At high temperature, we find scaling that is in accordance with the expected power law behavior of the Roberge-Weiss transition while a different behavior is found for $T\lesssim 170$ MeV.
\end{abstract}
  
\section{Introduction}
For the understanding and description of the fireball evolution in ultrarelativistic nucleus-nucleus collisions, knowledge of the QCD equation of state and the QCD phase diagram is essential. 
Lattice QCD calculations, which are very successful at nonzero temperature $T>0$ and vanishing net-baryon chemical potential $\mu_B=0$, are unfortunately hampered by the infamous sign problem at $\mu_B>0$. 
Many calculation strategies have been framed and refined in order to overcome the problem, but a working solution for calculations at finite $\mu_B$ has not been found so far. 
The most used methods are Taylor expansion at around $\mu_B=0$ \cite{Allton:2002zi} and calculations at purely imaginary chemical potential $\mu_B=i\mu^I_B$ \cite{DElia:2002tig, deForcrand:2002hgr} along with analytic continuation. 
Here we present a new method which is the combination of the above. We calculate Taylor expansion coefficients at various values of the imaginary chemical potential $\mu^I_B$ and base our analytic continuation to real $\mu_B$ on a multi-point Padé resummation \cite{Dimopoulos:2021vrk, Francesco:Lattice2022}. 

\begin{figure}
    \centering
    \includegraphics[width=0.6\textwidth]{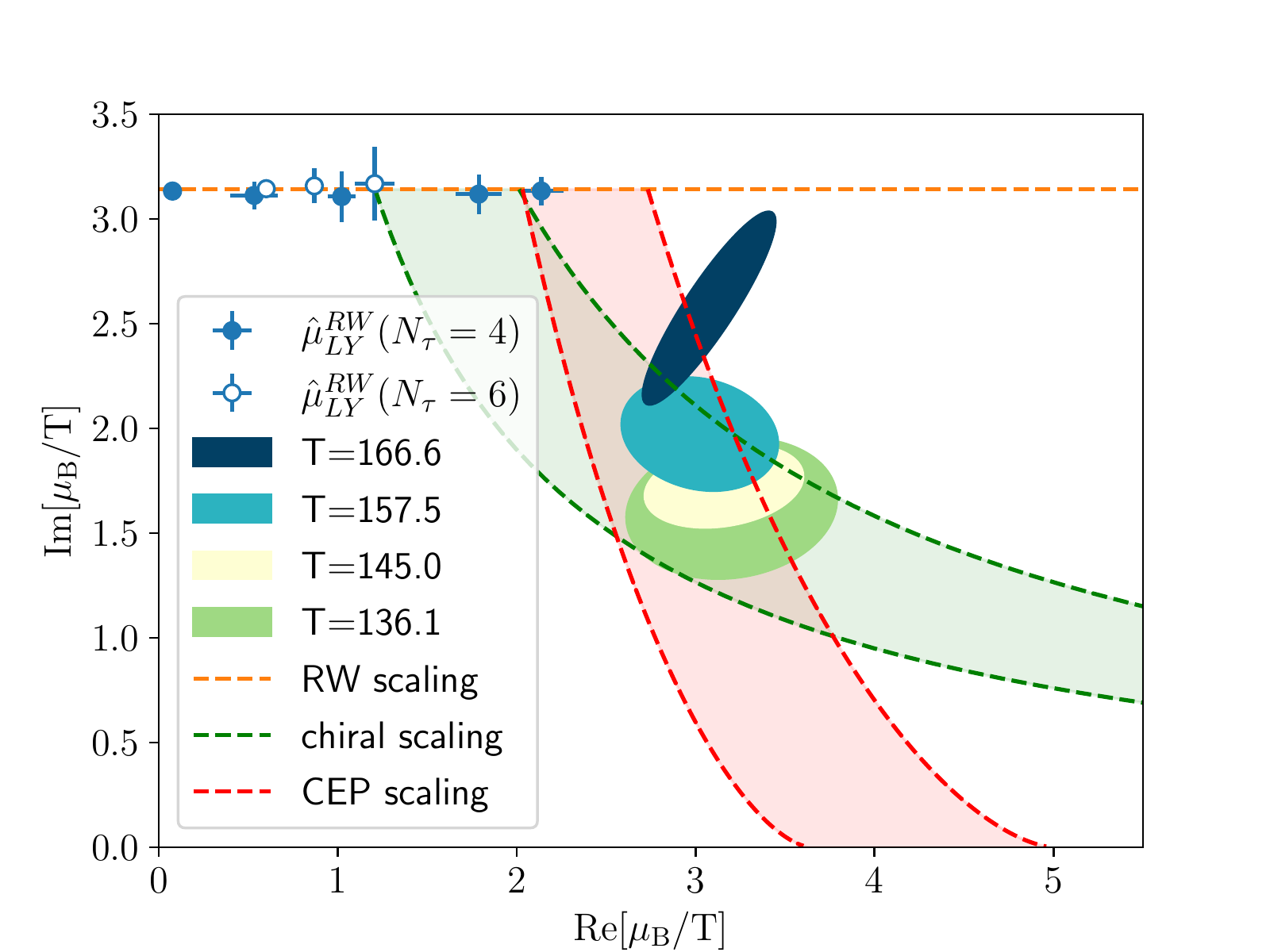}
    \caption{Expected scaling of the LYE singularity in the vicinity of the RW transition, the chiral transition and the QCD critical endpoint. Also shown are LYE singularities determined from lattice data: data points are from the temperature range $T\in[170-200]$ MeV ($N_\tau=4,6$), while confidence ellipses are indicated for lower temperatures ($N_\tau=6$). }
    \label{fig:overview}
\end{figure}

Padé approximants are rational functions and thus feature poles in the complex chemical potential plane.
The closest pole might be identified with the Lee-Yang edge (LYE) singularity in the vicinity of a critical point \cite{Yang:1952be}.
In terms of the scaling variable $z=t/h^{1/\beta\delta}$, where $t,h$ are the temperature and magnetic field-like variables, and $\beta, \delta$ are critical exponents, the LYE singularity has a universal position $z=z_c$ \cite{Connelly:2020gwa}.
We will make use of this universal position in the vicinity of the Roberge-Weiss (RW) transition, the chiral transition and the QCD critical end point to predict some non-universal constants and the generic scaling of the LYE in the QCD phase diagram. A schematic plot of the expected temperature scaling is shown in  Fig.~\ref{fig:overview}. With varying temperature, the LYE associated with a specific phase transition is expected to move within the respective band.
The ultimate goal is the prediction of the QCD critical end point, which is reached when the imaginary part of the LYE vanishes. 

\section{Lattice data and multi-point Padé analysis}
We use (2+1)-flavors of highly improved staggered quarks (HISQ) \cite{Follana:2006rc} for our simulations. 
The partition function can be written as 
\begin{equation}
    Z=\int\mathcal{D}U\; \det[M(m_l,i\mu_l^I)]^{2/4} 
    \; \det[M(m_s,i\mu_s^I)]^{1/4}\;e^{-S_G(U)}\,,
\end{equation}
where $M(m,i\mu^I)$ represents the fermion matrix of a HISQ flavor with mass $m$ and chemical potential $\mu=i\mu^I$. 
The first determinant represents the two degenerate light flavors (up and down quarks), the second one stands for the strange quarks. The respective quark masses have been tuned to yield physical meson masses in the vacuum. 
Details on the scale setting and lines of constant physics have been adopted from HotQCD \cite{Bollweg:2021vqf}. 
The chemical potentials are kept equal for simplicity ($\mu_l^I=\mu_s^I$) and are varied between $0$ and $i\pi/3$, which translates into $\mu_B^I\in[0,i\pi]$.
Configurations are generated using the $\simulat$ software \cite{Bollweg:2021cvl}.

\begin{figure}
    \centering
    \includegraphics[width=0.7\textwidth]{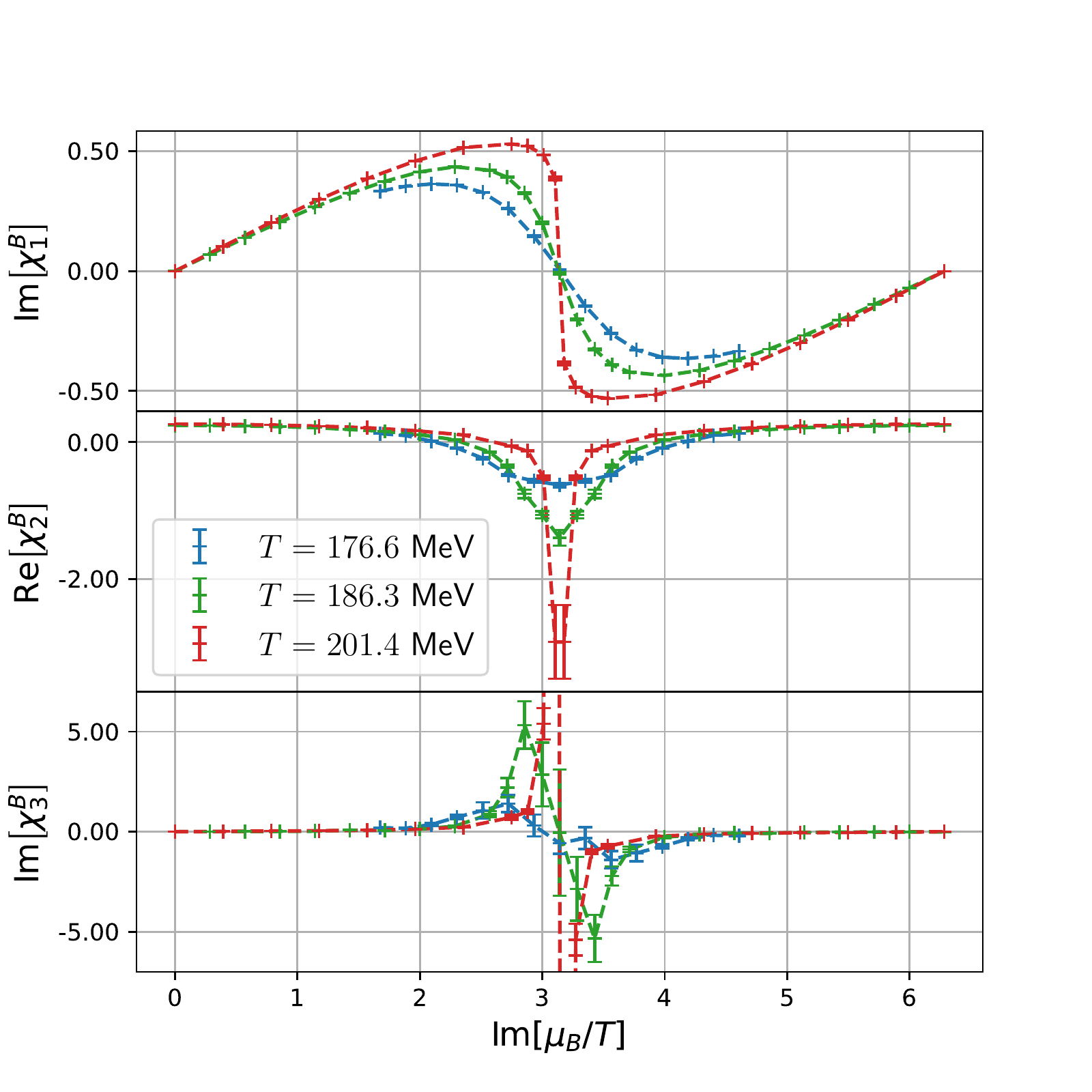}
    \caption{First three cumulants of the net-baryon number, for three different temperatures as function of an imaginary chemical potential. The data are obtained from calculations on $24^3\times 4$ lattices. }
    \label{fig:raw_data}
\end{figure}

Our observables are derivatives of the dimensionless pressure $p/T^4=\ln Z/(VT^3)$ with respect to $\hmu=\mu/T$. 
In particular we define cumulants of the net-baryon number $\chi_n^B$ as 
\begin{eqnarray}
    \chi_n^B(T,V,\mu_B)&=&\left(\frac{\partial}{\partial \hmu_B}\right)^n\frac{\ln Z(T,V,\mu_l,\mu_s)}{VT^3}\nonumber \\
    &=&\left(\frac{1}{3}\frac{\partial}{\partial \hmu_l}+\frac{1}{3}\frac{\partial}{\partial \hmu_s}\right)^n\frac{\ln Z(T,V,\mu_l,\mu_s)}{VT^3}\, .
\end{eqnarray}
Due to the symmetries of the partition function, its derivatives exhibit a distinct pattern. Besides the inherited periodicity in $i\mu_B^I$, we find that along the imaginary chemical potential axis, even derivatives are even functions of $i\hat\mu_B^I$ and purely real, and odd derivatives are odd functions and purely imaginary. The first three cumulants are shown in Fig.~\ref{fig:raw_data}.

We approximate the net-baryon density $\textrm{Im}\chi_1^B(\mu_B^I)$ by a rational function of the form 
\begin{equation}
    \textrm{Im}\chi_1^B(\mu_B^I)=\frac{P_m(\mu_B^I)}{1+Q_n(\mu_B^I)}\,
\end{equation}
where $P_m,Q_n$ are polynomials of order $m,n$ respectively. After multiplying with the denominator and demanding consistency with our data for $\chi_k^B$, with $k\in\{1,2,3\}$, we arrive at a linear system which we can solve \cite{Dimopoulos:2021vrk}. In this way, the expansion of the rational approximation agrees with our measured Taylor expansion at each of our simulation points. We also use different methods to obtain rational approximations, based on a generalized $\chi^2$-fitting approach and a variable transformation to the fugacity plan, which all  yield similar results \cite{Dimopoulos:2021vrk}. 
\section{Critical scaling}
\begin{figure}
    \centering
    \includegraphics[height=0.405\textwidth]{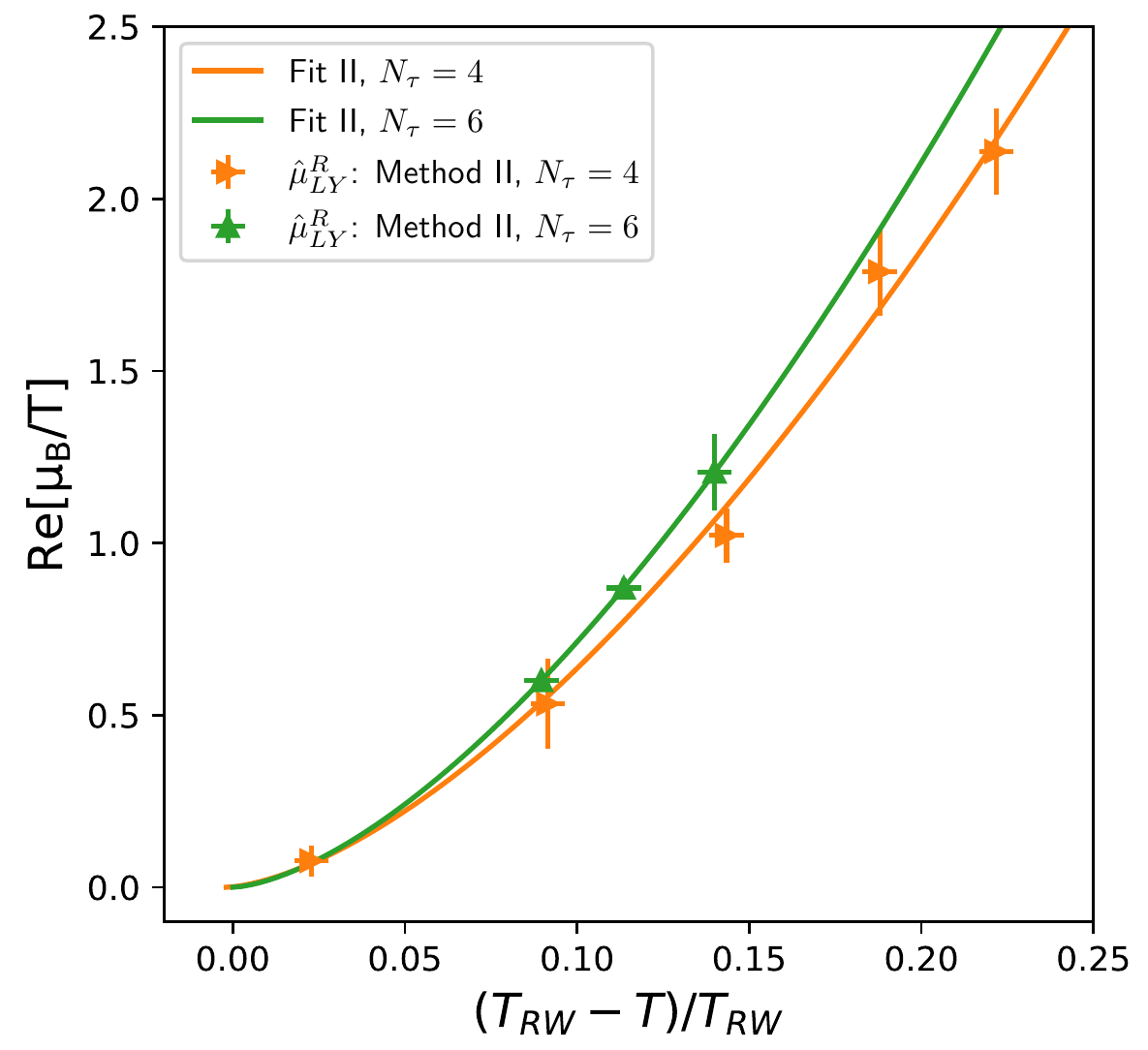}
    \includegraphics[height=0.397\textwidth]{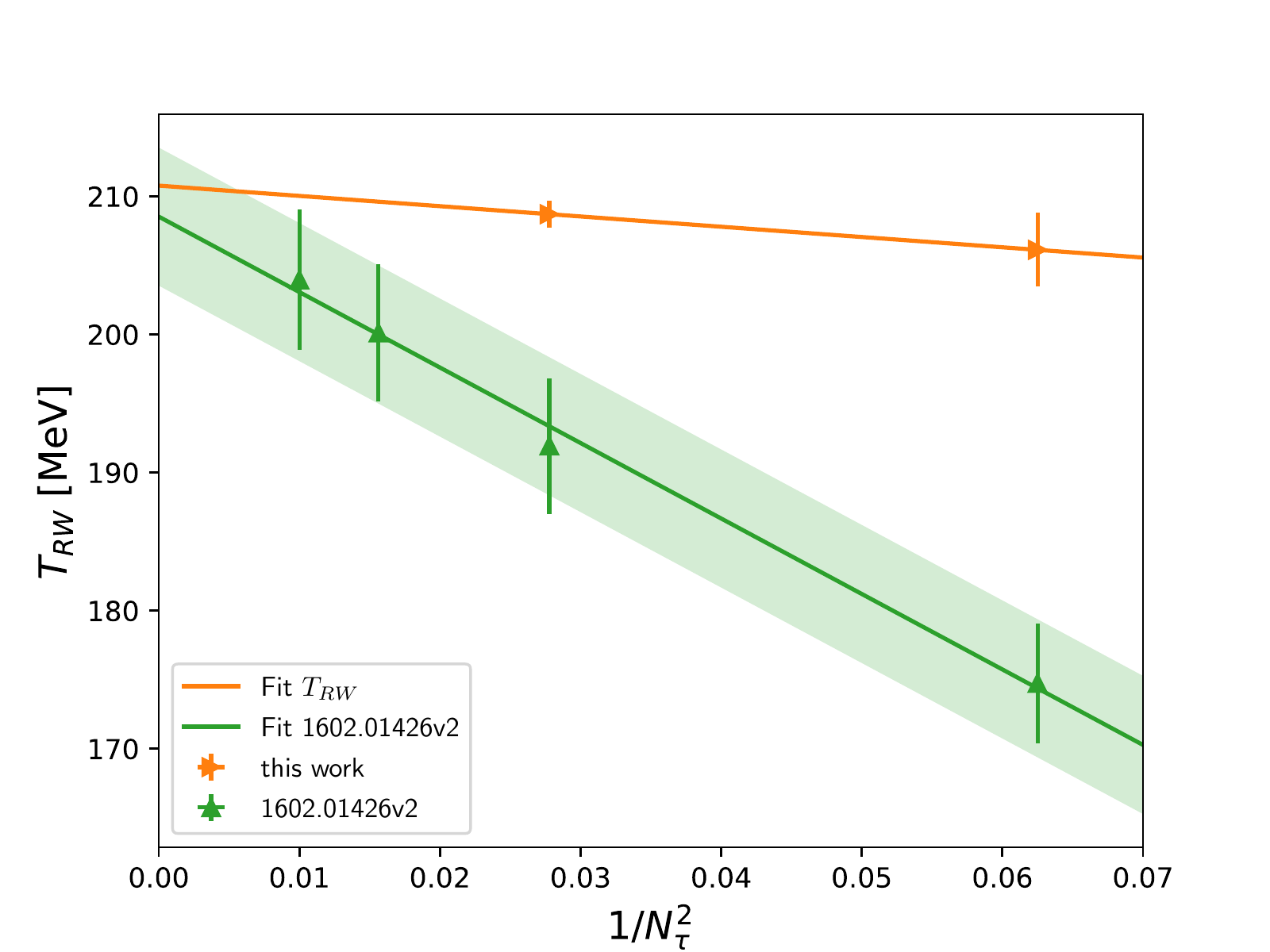}
    \caption{Left: Scaling of the real part of the LYE, with temperature. Lines are scaling fits to the $N_\tau=4$ and $6$ data, respectively. Right: Continuum estimate of the RW transition temperature $T_{RW}$, based on the scaling fists on the left. Also shown is a previous continuum extrapolation from the literature \cite{Bonati:2016pwz}.}
    \label{fig:RWscaling}
\end{figure}

Now we analyze the positions of the poles and roots of our rational approximations in the complex $\mu_B$ plane. We find for high temperatures ($T\in [170,200]$ MeV) that among many canceling pairs of roots and poles and the trivial roots at $\mu_B=ik\pi$, $k\in\mathbb{N}$, the poles and roots are alternating along the line $\mu_B=i\pi+\mu_B^R$. This is indeed the expected representation of the branch cut in the scaling function of the order parameter by a finite order $[m,n]$-Padé. We associate the closest pole of this structure with the LYE of the RW transition. The scaling of the real part of the LYE with temperature is shown in Fig.~\ref{fig:RWscaling} (left). 

When lowering $T$ from the RW transition temperature $T_{RW}$, where the LYE is located on the imaginary axis at $\mu_B=i\pi$, we find that the LYE moves into the complex plane and also obeys the expected scaling law \cite{Dimopoulos:2021vrk}, 
\begin{equation}
    \textrm{Re}\,\mu_{LY}\propto \left(\frac{T_{RW}-T}{T_{RW}}\right)^{\beta\delta}\,.
\end{equation}
From the fits to the $N_\tau=4,6$ data, using Z(2)-universal exponents $\beta\delta\approx 1.56$, we extract the RW transition temperature $T_{RW}$. The $N_\tau=4$ result is consistent with determinations from the peak positions of the Polyakov-Loop and chiral susceptibilities, using the same lattice action \cite{Cuteri:2022vwk}. A Preliminary continuum estimate, which is shown in Fig.~\ref{fig:RWscaling} (right), is in good agreement with a previous continuum extrapolation from \cite{Bonati:2016pwz} using the stout improved staggered action.

Interestingly, for temperatures below $T\lesssim 170$ MeV we find a qualitative change in the behavior of the LYE: the imaginary part becomes substantially different from $i\pi$ (see Fig.~\ref{fig:overview}) \cite{Kevin:Lattice2022}. 
It remains to be further analyzed whether our LYE data indicate chiral scaling and/or scaling which is related to the QCD critical point. 
The latter would give us a handle on the determination of its location \cite{Basar:2021hdf}. 
Some indications on chiral scaling in the temperature range of $T\in[170,180]$ MeV is found in the behavior of the Fourier coefficients of the baryon number density $\mathrm{Im} \chi_1^B(\mu_B^I)$ \cite{Christian:Lattice2022}. 

\section*{Acknowledgments}
This work is supported by Deutsche Forschungsgemeinschaft (DFG, German Research Foundation) through grant number 315477589, from the European Union under grant agreement No. H2020-MSCAITN-2018-813942 and the I.N.F.N. under the research project i.s. QCDLAT.
\printbibliography
\end{document}